\\
Title: Answer of the most important unsettled question of physical theory in 20$^{th}$ century
Author: Zhang Junhoa
Comments: 20 pages with 2 figures
Report-on:
Subj-class: Quantum mechanics
\\

Quantum mechanics take the sum of first finite order approximate solutions of time-dependent perturbation to substitute the exact solution. From the point of mathematics, it may be correct only in the convergent region of the time-dependent perturbation series. Where is the convergent region of this series? Quantum mechanics did not answer this problem. However it is relative to the question, can we use the Schrödinger equation to describe the transition processes? So it is the most important unsettling problem of physical theory.

We find out the time-dependent approximate solution $a_f^{(N)}(t)$ for arbitrary N and the exact solution $a_f(t) = \sum_{N=0}^{\infty} a_f^{(N)}(t)$. Then we can prove that: (1) In $|E_f - E_i| < |^{(S)}H'_{fi}|$ region, the series is divergent. The basic error of quantum mechanics is using the sum of the first finite orders approximate solutions to substitute the exact solution in this divergent region. It leads to an infinite error. So the Fermi's golden rule is not a mathematically reasonable inference of the. Schrödinger equation (2) The transiton probability per unit time deduced from the exact solution of Schrödinger equation cannot describe the transition processes.

We only give a prime discussion in this paper.
\\
## (1) The most important unsettled question of physical theory in 20$^{th}$ century

Quantum mechanics deems that, we can use the Schrödinger equation

$$i\hbar \frac{\partial \mathbf{y}}{\partial t} = \hat{H}(t)\mathbf{y} \tag{1}$$

to describe the variance of the system with time, where

$$\hat{H}(t) = \hat{H}_0 + \hat{H}'(t) = \hat{H}_0 + {}^{(S)}\hat{H}' g(t) \tag{2}$$

We may write the $\hat{H}'(t)$ as the product of time independent ${}^{(S)}\hat{H}'$ with a time dependent factor $g(t)$. Quantum mechanics only discusses two types $g(t)$, they are

$$g_1(t) = \begin{cases} 1, & if \quad 0 \leq t \leq \infty \\ 0, & if \quad t < 0 \end{cases} \tag{3}$$

or

$$g_2(t) = \begin{cases} \cos(\omega t + \delta) & \text{if} \quad 0 \leq t \leq \infty \\ 0 & \text{if} \quad t < 0 \end{cases} \quad (4)$$

The eigenvalues of $\hat{H}_0$ are discrete or continuous. We shall discuss for each case. In discrete case, according to quantum mechanics, suppose

$$\boldsymbol{y} = \sum_n a_n(t) \boldsymbol{f}_n \quad (5)$$

Then the Schrödinger equation may be written as

$$i\hbar \frac{da_f(t)}{dt} = \sum_k {}^{(S)}H_{fk}' g(t) \exp\left[\frac{i(E_f - E_k)t}{\hbar}\right] a_k(t) \quad (6)$$

The time-dependent perturbation theory supposes that

$$a_f(t) = a_f^{(0)}(t) + a_f^{(1)}(t) + a_f^{(2)}(t) + \cdots = \sum_{N=0}^{\infty} a_f^{(N)}(t) \quad (7)$$

and uses the approximate equation

$$i\hbar \frac{da_f^{(0)}(t)}{dt} = 0 \quad (8)$$

$$i\hbar \frac{da_f^{(N)}(t)}{dt} = \sum_k {}^{(S)}H'_{fk} g(t) \exp\left[\frac{i(E_f - E_k)t}{\hbar}\right] a_k^{(N-1)}(t) \quad \text{if} \quad N \geq 1 \quad (9)$$

to substitute the exact equation (6).

So far, people only find the few order approximate solutions of (7). We do not find out $a_k^{(N)}(t)$ for arbitrary $N$. The only way is to substitute the exact solution by the sum of first finite order approximate solutions

$$a_f(t) \rightarrow \sum_{N=0}^{M} a_f^{(N)}(t) \quad (10)$$

and to use this sum to describe the transition processes.

The problem is that, whether the $\sum_{N=M+1}^{\infty} a_f^{(N)}(t)$ may be abandoned? From the point of physics, if the abandoned part is large then the experimental error, then we cannot obtain any conclusion from the comparison between the sum of first finite order approximate solutions with experimental results. In this case, the words "Schrödinger equation may describe the transition processes" would be meaningless. From the point of mathematics, $a_f(t)$ is the function of energy $E_f$. Only in the convergent range of series, for $\boldsymbol{e} > 0$, we can find a number M, that the relation

$$\left| a_f(t) - \sum_{N=0}^{M} a_f^{(N)}(t) \right| < \varepsilon \qquad (11)$$

is satisfied. What is $a_f^{(N)}(t)$ for arbitrary $N$? What is the exact solution $a_f(t) = \sum_{N=0}^{\infty} a_f^{(N)}(t)$?

In which range is the series divergent? These question are relative to that, does the equation of quantum mechanics can describe the transition processes? So it is the most important unsettled question of physical theory.

## (2) Discussion in the case in which the eigenvalues of $\hat{H}_0$ are discrete and the time factor is $g_1(t)$

The equations (6) and (9) contain $\sum_k$, it means that we discuss in the discrete case. We shall solve the Schrödinger equation and its approximate equations in the range $0 \leq t$.

## (2-1) Expression of $a_f^{(N)}(t)$ for arbitrary $N$

The key of the problem is to find $a_f^{(N)}(t)$ for arbitrary $N$. We shall deduce $a_f^{(N)}(t)$ in other paper [8]. In this paper we point that the expression of $a_f^{(N)}(t)$ is

$$a_f^{(N)}(t) = \sum_{\lambda} \left[ b_{fi}^{(N)}(\lambda) + \sum_{n=1}^{N} \frac{1}{n!} \left( \frac{-it}{\hbar} \right)^n \sum_{M=n}^{N} b_{fi}^{(N-M)}(\lambda) \theta^{(M)}(\lambda, n) \right] \times \exp\left[ \frac{i(E_f - \lambda^{(0)})t}{\hbar} \right] \qquad (12)$$

However we prove that it is the solution of (8) and (9) in appendix A. Each term in (12) is the product of a polynomial of $t$ with $\exp[i(E_f - \lambda^{(0)})t/\hbar]$. $b_{fi}^{(N)}$, $\theta^{(M)}$ are the time-dependent coefficient. The stationary-state equation is

$$\lambda c_f(\lambda) = \sum_k (E_k \delta_{fk} + {}^{(S)}H'_{fk}) c_k(\lambda) \qquad (13)$$

The orthonormality condition of $c_l(\boldsymbol{l})$ is

$$\sum_{\lambda} c_f(\lambda) c_i(\lambda)^* = \delta_{fi} \qquad (14)$$

Suppose

$$\lambda = \sum_{N=0}^{\infty} \lambda^{(N)} \qquad (15)$$

$$c_f(\lambda) = \sum_{N=0}^{\infty} c_f^{(N)}(\lambda) \tag{16}$$

The perturbed approximate stationary-state equations are

$$\boldsymbol{\lambda}^{(0)} c_f^{(0)}(\boldsymbol{\lambda}) - \sum_k E_f \boldsymbol{d}_{fk} c_k^{(0)}(\boldsymbol{\lambda}) = 0$$

$$(\boldsymbol{\lambda}^{(0)} - E_f) c_f^{(N)}(\boldsymbol{\lambda}) + \sum_{M=1}^{N} \boldsymbol{\lambda}^{(M)} c_f^{(N-M)}(\boldsymbol{\lambda}) = \sum_k {}^{(S)}H'_{fk} c_k^{(N-1)}(\boldsymbol{\lambda}) \qquad N > 0 \tag{17}$$

where $c_f^{(N)}(\lambda)$ satisfies the orthonomality condition

$$\sum_\lambda c_f^{(0)}(\lambda) c_i^{(0)}(\lambda)^* = \delta_{fi}$$

$$\sum_{\boldsymbol{\lambda}} \sum_{L=0}^{N} c_f^{(L)}(\boldsymbol{\lambda}) c_i^{(N-L)}(\boldsymbol{\lambda})^* = 0, \qquad if \quad N > 0 \tag{18}$$

$b_{fi}^{(N)}$, $\boldsymbol{q}^{(M)}(\boldsymbol{\lambda},n)$ are the functions of $\boldsymbol{\lambda}^{(N)}$, $c_k^{(N)}(\boldsymbol{\lambda})$.

### (2-2) The expression of the exact solution

According to (7), the exact solution is

$$a_f(t) = \sum_{N=0}^{\infty} a_f^{(N)}(t)$$

$$= \sum_\lambda \sum_{N=0}^{\infty} \left[ b_{fi}^{(N)}(\lambda) + \sum_{n=1}^{N} \frac{1}{n!} \left( \frac{-it}{\hbar} \right)^n \sum_{M=n}^{N} b_{fi}^{(N-M)}(\lambda) \theta^{(M)}(\lambda, n) \right] \times \exp\left[ \frac{i(E_f - \lambda^{(0)})t}{\hbar} \right]$$

(19)

Calculating this expression, we may obtain

$$a_f(t) = \sum_\lambda c_f(\lambda) c_i(\lambda)^* \exp\left[ \frac{i(E_f - \lambda)t}{\hbar} \right] \tag{20}$$

You can find the calculating process in appendix B. The opposite process is to expand $\lambda$, $c_f(\lambda)c_i(\lambda)^*$ presented in (20) by using the stationary-state perturbed approximation, we may obtain the different order of time-dependent perturbed approximate solution $a_f^{(N)}(t)$.

Finally we test that $a_f(t)$ is the exact solution of the Schrödinger equation in appendix C.

### (2-3) The error of time-dependent perturbed approximation

Now we affirm three points:
(i) (20) is satisfied the Schrödinger equation and the special initial condition;

(ii) Expanding the coefficient $\lambda$, $c_f(\lambda)$ presented in $a_f(t)$ according stationary-state

perturbed approximation, and classifying according the approximate order we get $a_f^{(N)}(t)$;

(iii) $a_f^{(N)}(t)$ is satisfied the time-dependent perturbed approximate equation and special initial condition.

They are the basic to discuss the core of unsettled question—in which range is the series divergent?

The exact solution $a_f(t)$ must be expressed by the solution of stationary state equation $1$, $c_f(1)$. Substituting the stationary state perturbed approximate solution into $1$, $c_f(1)$ presented in the exact solution, we shall obtain the different order approximations of the time–dependent perturbed method $a_f^{(N)}(t)$. Therefore, the divergent intervals of the time-dependent perturbed approach and the one of the stationary state perturbed approach are identical. Quantum mechanics pointed out that the divergent interval of the stationary state perturbed series is

$$|E_f - E_i| < |{}^{(S)}H'_{fi}| \tag{21}$$

So it must be the divergent interval of the time-dependent perturbed series. It means that, for each $e > 0$, no matter how much $M$ order approach solutions we take into account, the relation

$$\left| a_f(t) - \sum_{N=0}^{M} a_f^{(N)}(t) \right| < \varepsilon \tag{22}$$

does not satisfy in the divergent interval. We cannot use the sum of first finite order of time-dependent perturbed solutions to substitute the exact solution. **The error of time-dependent perturbed method is using $\sum_{N=0}^{M} a_f^{(N)}(t)$ to substitute $a_f(t)$ for calculating the transition probability in the divergent range $|E_f - E_i| < |{}^{(S)}H'_{fi}|$.** Let us examine the first-order approximation of the time-dependent perturbed approximation

$$a_f^{(1)}(t)\Big|_{f \neq i} = \sum_l \left[ c_f^{(1)}(\lambda^{(0)} = E_l) c_i^{(0)}(\lambda^{(0)} = E_l)^* + c_f^{(0)}(\lambda^{(0)} = E_l) c_i^{(1)}(\lambda^{(0)} = E_l)^* \right]$$

$$\times \exp\left[\frac{i(E_f - \lambda^{(0)})t}{\hbar}\right]$$

$$= c_f^{(1)}(\lambda^{(0)} = E_i) \exp\left[\frac{i(E_f - E_i)t}{\hbar}\right] + c_i^{(1)}(\lambda^{(0)} = E_f)^*$$

$$= \frac{{}^{(S)}H'_{fi}}{E_i - E_f}\left\{\exp\left[\frac{i(E_f - E_i)t}{\hbar}\right] - 1\right\}$$

(23)

$c_i(\mathbf{l})$ is the $\mathbf{l}$-component of an unit vector. Therefore

$$\left|c_f^{(1)}(\lambda^{(0)} = E_i)\right| = \left|\frac{{}^{(S)}H'_{fi}}{E_i - E_f}\right| \leq 1 \tag{24}$$

and (23) can only be used in the region $\left|{}^{(S)}H'_{fi}\right| \leq \left|E_i - E_f\right|$.

From quantum mechanics, we have

$$w_{i\to f}^{(1)} = \sum 4\left|{}^{(S)}H'_{fi}\right|^2 \frac{\sin^2\left[(E_f - E_i)t/(2\hbar)\right]}{(E_f - E_i)^2} \rho(E_f)\Delta E_f \tag{25}$$

$$\to \int 4\left|{}^{(S)}H'_{fi}\right|^2 \frac{\sin^2\left[(E_f - E_i)t/(2\hbar)\right]}{(E_f - E_i)^2} \rho(E_f)dE_f$$

$$= \left(\frac{2\pi t}{\hbar}\right)\left|{}^{(S)}H'_{fi}\right|^2 \rho(E_f). \tag{26}$$

There are two questions. First, the premise of this section is that we discuss in the discrete case. Quantum mechanics changes to discuss the continuous case and simply substitutes (25) with (26). Is this reasonable? We shall discuss this question in the next section. Second, the transition probability deduced from the perturbed approximation is correct only in the region $\left|{}^{(S)}H'_{fi}\right| < \left|E_f - E_i\right|$; the fault is in $\left|E_f - E_i\right| < \left|{}^{(S)}H'_{fi}\right|$.

Quantum mechanics discuss the relation between $\sin^2[\Delta E t/(2\hbar)]/(\Delta E)^2$ with $\Delta E = E_f - E_i$. And pointed out there is a peak of function in $\frac{|\Delta E_1|t}{2\hbar} = \pi$, this peak of function is just deduced from the error factor $(E_i - E_f)^{-2}$. If $t \to \infty$, the high of peak tends to infinite, and the breadth of the peak tends to zero, the discussed function becomes $\delta$ function. However if $t = \dfrac{2\pi\hbar}{\left|{}^{(S)}H'_{fi}\right|}$, the whole main peak of $\delta$ function is in the divergent region of the perturbed approximate series. The problem is that it is under these two conditions, $t \geq \dfrac{2\pi\hbar}{\left|{}^{(S)}H'_{fi}\right|}$, and $|\Delta E| \leq \left|{}^{(S)}H'_{fi}\right|$, that quantum mechanics obtains

$$w_{i \to f} = \dfrac{2\pi}{\hbar}\left|{}^{(S)}H'_{fi}\right|^2 \rho(E_f) \tag{27}$$

from (26). Fermi called it as the golden rule. Of course, if this formula is correct, then someone can find a suitable $\left|{}^{(S)}H'_{fi}\right|$ to be identical with the experimental value $w_{i \to f}$, but mathematically Fermi's golden rule is not a reasonable conclusion from the Schrödinger equation.

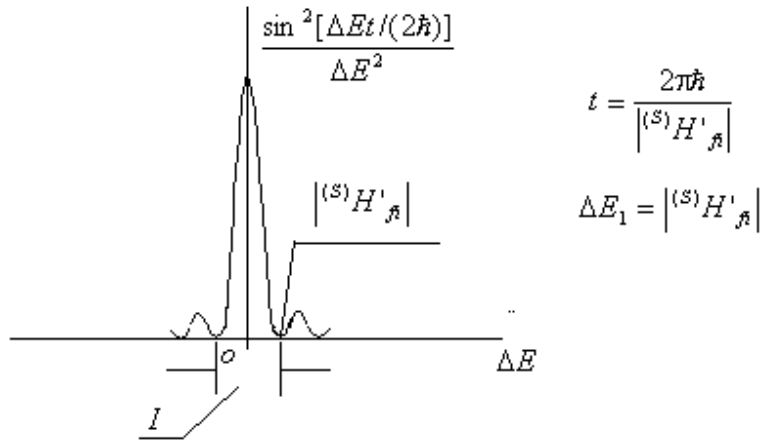

Fig.1 The relation between $\sin^2[\Delta E t/(2\hbar)]/(\Delta E)^2$ with $\Delta E = E_f - E_i$, where I is the divergent region of the time-dependent perturbed approximate series.

**(2-4) Transition probability per unit time deduced from the exact solution**

According to quantum mechanics, the transition probability per unit time is defined as

$$\begin{aligned} w_{i \to f} &= \lim_{t \to \infty} \sum_{E_l = E_f - \Delta E/2}^{E_f + \Delta E/2} \Delta E_l\, \mathbf{r}(E_l) a_l(t)^* a_l(t)/t \\ &= \lim_{t \to \infty} \sum_{E_l = E_f - \Delta E/2}^{E_f + \Delta E/2} \Delta E_l\, \mathbf{r}(E_l) \sum_{\mathbf{n}} \sum_{\mathbf{l}} c_l(\mathbf{l})^* c_i(\mathbf{l}) c_l(\mathbf{n}) c_i(\mathbf{n})^* t^{-1} \\ &\quad \times \exp\left[\dfrac{i(\mathbf{l}-\mathbf{n})t}{\hbar}\right] \end{aligned} \tag{28}$$

Note that

$$\sum_{\mathbf{l}} c_l(\mathbf{l})^* c_l(\mathbf{l}) = 1 \tag{29}$$

It means that $c_l(\lambda)$ is the $\lambda$-component of one unit vector, and $c_i(\lambda)\exp(i\lambda t/\hbar)$ is the $\lambda$-component of another unit vector. The expression

$$\sum_{\lambda} c_l(\lambda)^* c_i(\lambda)\exp(i\lambda t/\hbar)$$

is the product of two unit vectors, so

$$\left|\sum_{\mathbf{l}} c_l(\mathbf{l})^* c_i(\mathbf{l})\exp(i\mathbf{l}t/\hbar)\right| \leq 1 \tag{30}$$

Finally we obtain

$$w_{i \to f} \leq \lim_{t \to \infty} \sum_{E_l = E_f - \Delta E/2}^{E_f + \Delta E/2} \dfrac{\rho(E_l)\Delta E_l}{t} = 0 \tag{31}$$

This is the result deduced from the exact solution of Schrödinger equation (1) in the $g_1(t)$ case.

This result cannot be used to describe the transition processes.

We summarize the discussion in the discrete case:

(i) The substance of the time-dependent perturbed approximation is to expand $\lambda$, $c_k(\lambda)$ presenting the expression of $a_f(t)$ according to the stationary-state perturbed approach..

(ii) The convergence interval of the stationary-state perturbed approach is $\left|{}^{(S)}H'_{fi}\right| \leq \left|E_f - E_i\right|$, which is also the convergence interval of the time-dependent perturbed approach..

(iii) The sum of the first finite order perturbed approximate solutions may be used to substitute the exact solution only in the convergence interval. The fundamental error of the time-dependent perturbed approach is using the sum of the first finite order approximate solutions to calculate the transition probability in the interval $\left|{}^{(S)}H'_{fi}\right| > \left|E_f - E_i\right|$. Mathematically Fermi's golden rule is not a reasonable conclusion from the Schrödinger equation in the discrete case.

(iv) People use the sum of first finite order perturbed approximate solutions to substitute the exact solution. If only caused by that they did not find out the approximate solution $a_f^{(N)}(t)$ for arbitrary N. When we find out $a_f^{(N)}(t)$ and $a_f(t) = \sum_{N=0}^{\infty} a_f^{(N)}(t)$, then we have not any reason to use the sum of first finite order perturbed approximate solutions to substitute the exact solution. The pity is that, the transition probability per unit time deduced from the exact solution of the Schrödinger equation is zero in the discrete case, it cannot be used to describe the transition process.

## (3) Discussion in the case in which the eigenvalues of $\hat{H}_0$ are continuous and the time factor is $g_1(t)$

In the continuous case, the discrete variable $k$ must change to continuous variable $E_k$, the summation must change to the integration, and $\delta_{fi}$ presented in the discrete case must change to $\delta(E_f - E_i)$, the others are same as in the discrete case. This kind of variation cannot give the difference result.

### (3-1) Exact solution

In the continuous case the Schrödinger equation has the form

$$i\hbar \frac{da_f(t)}{dt} = \int dE_k \, {}^{(S)}H'_{fk} \exp\left[\frac{i(E_f - E_k)t}{\hbar}\right] a_k(t) \tag{32}$$

The initial condition can be expressed in the form

$$a_f(t=0) = A\pmb{\delta}(E_f - E_i) \tag{33}$$

If the initial condition takes the form of (33), then at $t = 0$ the probability of the system in the $(E_i - \Delta E/2, \ E_i + \Delta E/2)$ region is

$$\int_{E_i-\Delta E/2}^{E_i+\Delta E/2} a_l(t=0)^* a_l(t=0) dE_l$$

$$= |A|^2 \int_{E_i-\Delta E/2}^{E_i+\Delta E/2} \pmb{\delta}(E_l - E_i)\pmb{\delta}(E_l - E_i) dE_l = |A|^2 \pmb{\delta}(0) \tag{34}$$

We can prove that the exact solution of the Schrödinger equation (32) and of the initial condition is

$$a_f(t) = A\int d\pmb{\lambda}\, c_f(\pmb{\lambda}) c_i(\pmb{\lambda})^* \exp\left[\frac{i(E_f - \pmb{\lambda})t}{\hbar}\right] \tag{35}$$

where $\pmb{\lambda}$, $c_f(\pmb{\lambda})$ are the eigenvalue and eigenfunction of the stationary-state equation

$$\int dE_k [E_f \delta(E_f - E_k) + {}^{(S)}H'_{fk}] c_k(\lambda) = \lambda c_f(\lambda) \tag{36}$$

respectively.

At the same time, the eigenfunctions satisfy the orthonormality conditions

$$\int d\lambda\, c_f(\lambda) c_i(\lambda)^* = \delta(E_f - E_i) \tag{37}$$

$$\int dE_j\, c_j(\lambda) c_j(\nu)^* = \delta(\lambda - \nu) \tag{38}$$

**(3-2) Equivalence of the time-dependent perturbed approximation and the stationary-state perturbed approximation**

According to the stationary-state perturbed approximation method, suppose that

$$\pmb{\lambda} = \sum_{N=0}^{\infty} \pmb{\lambda}^{(N)} \qquad c_j(\pmb{\lambda}) = \sum_{N=0}^{\infty} c_j^{(N)}(\pmb{\lambda}) \tag{39}$$

where $\lambda^{(N)}$, $c_j^{(N)}(\lambda)$ satisfy a set of equations

$$(\lambda^{(0)} - E_f) c_f^{(0)} = 0,$$

$$(\lambda^{(0)} - E_f) c_f^{(N)}(\lambda) + \sum_{L=1}^{N} \lambda^{(L)} c_f^{(N-L)}(\lambda) = \int dE_k\, {}^{(S)}H'_{fk}\, c_k^{(N-1)}, \tag{40}$$

and satisfy the orthonormality condition

$$\int d\lambda\, c_f^{(0)}(\lambda) c_i^{(0)}(\lambda)^* = \delta(E_f - E_i) \tag{41}$$

$$\int d\lambda \sum_{L=0}^{N} c_f^{(L)}(\lambda) c_i^{(N-L)}(\lambda)^* = 0. \tag{42}$$

Inserting the stationary-state perturbed approach (39) into (35), and decomposing them according to the approximate order, we obtain

$$a_f(t) = \sum_{N=0}^{\infty} a_f^{(N)}(t) \tag{43}$$

where

$$a_f^{(N)}(t) = A \int d\mathbf{I} \left[ b_{fi}^{(N)}(\mathbf{I}) + \sum_{n=1}^{N} \frac{1}{n!} \left( \frac{-it}{\hbar} \right)^n \sum_{L=n}^{N} b_{fi}^{(N-L)}(\mathbf{I}) \mathbf{q}^{(L)}(\mathbf{I},n) \right] \exp\left[ \frac{i(E_f - \mathbf{I}^{(0)})t}{\hbar} \right] \tag{44}$$

The relations between $b_{fi}^{(N)}(\lambda)$, $\theta^{(L)}(\lambda,n)$ with $\lambda^{(N)}$, $c_j^{(N)}(\lambda)$ are same as in the discrete case.

$a_f^{(N)}(t)$ are the solution of the time-dependent perturbed approximation equations

$$\begin{aligned} i\hbar \frac{da_f^{(0)}(t)}{dt} &= 0 \\ i\hbar \frac{da_f^{(N)}(t)}{dt} &= \int dE_k \, {}^{(S)}H'_{fk} \exp\left[ \frac{i(E_f - E_k)t}{\hbar} \right] a_k^{(N-1)} \end{aligned} \tag{45}$$

The approximate solutions satisfy the initial conditions

$$a_f^{(0)}(t=0) = A\mathbf{d}(E_f - E_i) \tag{46}$$

$$a_f^{(N)}(t=0) = 0 \qquad N > 0 \tag{47}$$

The time-dependent perturbed approximation is equivalent to the stationary-state perturbed approximation in this case.

**(3-3) Error of deduction of transition probability from the time-dependent perturbed approach**

The zeroth and first-order stationary-state perturbed approximations are

$$c_f^{(0)}(\mathbf{I}^{(0)} = E_i) = \mathbf{d}(E_i - E_f) \tag{48}$$

$$c_f^{(1)}(\lambda^{(0)} = E_i) = \frac{{}^{(S)}H'_{fi}}{E_i - E_f} \tag{49}$$

In the neighborhood of the main peak of $\mathbf{d}$ function, we have

$$\frac{\left| c_f^{(1)}(\lambda^{(0)} = E_i) \right|}{\left| c_f^{(0)}(\lambda^{(0)} = E_i) \right|_{\max}} = \frac{\left| {}^{(S)}H'_{fi} \right|}{\left| E_i - E_f \right|} \cdot \frac{\hbar\pi}{t} > 1 \tag{50}$$

where $|\Delta E|t \to 0$, in this case. So we cannot use the sum of first finite order perturbed approximate solution to substitute the exact solution.

### (3-4) Deduction of the transition probability from the exact solution

The transition probability in $(0, t)$ is a relative probability. It is defined as

$$W(t) = \frac{\text{the probability at } t \text{ in } (E_f - \Delta E/2, \ E_f + \Delta E/2)}{\text{the probability at } t = o \text{ in all energy regions}}$$

$$= \frac{\int_{E_f - \Delta E/2}^{E_f + \Delta E/2} a_l^*(t) a_l(t) \rho(E_l) dE_l}{\int_{\text{all energy}} a_l^*(0) a_l(0) dE_l} \tag{51}$$

In this equation the denominator is

$$\int_{\text{all energy}} a_l(t=0)^* a_l(t=0) dE_l$$

$$= |A|^2 \int_{E_i - \Delta E/2}^{E_i + \Delta E/2} \delta(E_l - E_i) \delta(E_l - E_i) dE_l = |A|^2 \delta(0); \tag{52}$$

the numerator is

$$\int_{E_f - \Delta E/2}^{E_f + \Delta E/2} \rho(E_l) dE_l a_l^*(t) a_l(t)$$

$$= |A|^2 \int_{E_f - \Delta E/2}^{E_f + \Delta E/2} \rho(E_l) dE_l \int c_l(\lambda) c_i^*(\lambda) \exp\left[\frac{-i\lambda t}{\hbar}\right] d\lambda \int c_l^*(\nu) c_i(\nu) \exp\left[\frac{i\nu t}{\hbar}\right] d\nu \tag{53}$$

In the continuous case, $c_l(\mathbf{l})$ satisfies the orthonormality condition (37). Then $c_l(\mathbf{l})$ is the $\mathbf{l}$-component of one normality vector, and $c_i^*(\mathbf{l}) \exp(-i\mathbf{l}t/\hbar)$ is the $\mathbf{l}$-component of another normality vector. The expression

$$\int d\lambda c_l(\lambda) c_i^*(\lambda) \exp\left(\frac{-i\lambda t}{\hbar}\right)$$

is the product of two normality vectors, so

$$\left| \int d\lambda c_l(\lambda) c_i^*(\lambda) \exp\left(\frac{-i\lambda t}{\hbar}\right) \right| \leq \delta(E_l - E_i) \tag{54}$$

The transition probability per unit time is

$$w_{i \to f} = \lim_{t \to \infty} \frac{W(t)}{t}$$

$$\leq \lim_{t \to \infty} \left( \frac{1}{t \mathbf{d}(0)} \right) \int_{E_f - \Delta E/2}^{E_f + \Delta E/2} \mathbf{r}(E_l) dE_l \mathbf{d}(E_l - E_i) \mathbf{d}(E_l - E_i) = \lim_{t \to \infty} \frac{\mathbf{r}(E_f)}{t} = 0. \tag{55}$$

In the continuous case the transition probability per unit time deduced from the exact solution of the Schrödinger equation is still zero.

## (4) Discussion in the case in which the time factor is $g_2(t)$ and the eigenvalues of $\hat{H}_0$ are discrete

Quantum mechanics uses

$$\hat{H}'(t) = g_2(t)\hat{B} = \hat{B}\cos(\boldsymbol{\omega} t + \boldsymbol{\delta}), \quad 0 < t \tag{56}$$

to describe the electron transition from one energy level to another due to the impact of the electromagnetic field. In general, only the discrete is discussed. The Schrödinger equation has the form

$$i\hbar \frac{da_f(t)}{dt} = \sum_k \left\{ F_{fk} \exp\left[\frac{i(E_f + \hbar\omega - E_k)t}{\hbar}\right] + F_{fk}^* \exp\left[\frac{i(E_f - \hbar\omega - E_k)t}{\hbar}\right] \right\} a_k(t) \tag{57}$$

where

$$F_{fk} = B_{fk} e^{i\boldsymbol{\delta}} / 2 \tag{58}$$

Suppose

$$a_f(t) = \sum_{L=-\infty}^{\infty} A_{fL}(t) \tag{59}$$

where $A_{fL}(t)$ is the probability amplitude of the state in which there is an electron in $E_f$ and $L$ photons of energy $\hbar\boldsymbol{\omega}$. If at $t = 0$, the system is in the state $(E_i, L_0\hbar\omega)$, then

$$A_{fL}(t = 0) = \boldsymbol{\delta}_{fi}\boldsymbol{\delta}_{LL_0} \tag{60}$$

According to (57) and (60), we get

$$A_{fL}(t) = \sum_{\boldsymbol{\mu}} c_{fL}(\boldsymbol{\mu}) c_{iL_0}^*(\boldsymbol{\mu}) \exp\left[\frac{i(E_f + L\boldsymbol{\omega}\hbar - \boldsymbol{\mu})t}{\hbar}\right] \tag{61}$$

which $\mu$, $c_{fL}(\mu)$ are the solution of the stationary-state equation

$$(\mu - E_f - L\hbar\omega)c_{fL}(\mu) = \sum_k \left[F_{fk} c_{k,L-1}(\mu) + F_{fk}^* c_{k,L+1}(\mu)\right] \tag{62}$$

Suppose

$$K_{fL,kL'} = \begin{cases} F_{fk} & \text{if } L' = L - 1 \\ F_{fk}^* & \text{if } L' = L + 1 \\ 0 & \text{if } L' \neq L \pm 1 \end{cases} \tag{63}$$

then (62) may be rewritten in the form

$$\nu c_{fL}(\nu) = \sum_k \sum_{L'} \left[ (E_f + L\omega\hbar)\delta_{fk}\delta_{LL'} + K_{fL,kL'} \right] c_{kL'}(\nu) \tag{64}$$

At the same time, $c_{fL}(n)$ meets the orthonormality conditions of the eigenfunctions:

$$\sum_\nu c_{fL}(\nu)c^*_{iL'}(\nu) = \delta_{fi}\delta_{LL'} \tag{65}$$

$$\sum_j \sum_L c_{jL}(\lambda)c^*_{jL}(\nu) = \delta_{\lambda\nu} \tag{66}$$

Obviously, the exact solutions solved under the cases $g_1(t)$ and $g_2(t)$ have the identical mathematical structure. The results in Sec. 2 are still applicable in this case.

**Conclusion**

(i)  In the cases in which the time factors are $g_2(t)$ or $g_2(t)$, the eigenvalues of $\hat{H}_0$ are discrete or continuous, the process to deduce the transition probability per unit time from time-dependent perturbed approximation contains a basic mathematic mistake. It is using the sum of first finite order approximate solutions to substitute the exact solution for calculating the transition probability in the divergent range $\left|{}^{(S)}H'_{fi}\right| > \left|E_f - E_i\right|$. Fermi's golden rule is not the mathematically reasonable deductive inference from the Schrödinger equation.

(ii) The transition probability per unit time deduced from the exact solution of the Schrödinger equation cannot be used to describe time-dependent processes.

**(5) Other equations in quantum mechanics**

There are other equations in quantum mechanics, such as the unitary matrix equation

$$U_{fi}(t,t_0) = \delta_{fi} + \frac{1}{i\hbar}\int_{t_0}^t dt' \sum_k {}^{(I)}H'_{fk}(t)U_{ki}(t',t_0) \tag{67}$$

and the Dyson's equation (method of Green's function)

$$\Psi(\mathbf{r},t) = \iint d\mathbf{r}' dt' G_0(\mathbf{rr}'tt') {}^{(S)}H'(\mathbf{r}')\Psi(\mathbf{r}',t') \\ + \Phi_i(\mathbf{r})\exp(-iE_i t/\hbar) \tag{68}$$

The unitary matrix equation is suited to extend to quantization of field. We have discussed these two equations. The result is same as the Schrödinger equation. The exact solution cannot be used to describe the time dependent processes.

The basic task of quantum mechanics is to describe the stationary motion of the system and the time dependent processes. If the exact solution cannot be used to describe the time dependent processes, then we must reconsider the basic point and the basic picture of quantum mechanics.

# Appendix A The time-dependent perturbed approximate solution $a_f^{(N)}(t)$ is (12) for arbitrary N

Now let us prove that the time-dependent perturbed approximate solution is

$$a_f^{(N)}(t) = \sum_\lambda \left[ b_{fi}^{(N)}(\lambda) + \sum_{n=1}^{N} \frac{1}{n!}\left(\frac{-it}{\hbar}\right)^n \sum_{M=n}^{N} b_{fi}^{(N-M)}(\lambda)\theta^{(M)}(\lambda,n) \right] \times \exp\left[\frac{i(E_f - \lambda^{(0)})t}{\hbar}\right]$$

(12)

by the substitutive method, where

$$b_{fi}^{(N)}(\mathbf{l}) = \begin{cases} \sum_{L=0}^{N} c_f^{(L)}(\mathbf{l}) c_i^{(N-L)}(\mathbf{l})^* & \text{if } N \geq 0 \\ 0 & \text{if } N < 0 \end{cases} \quad (A1)$$

$$\mathbf{q}^{(M)}(\mathbf{l},1) = \mathbf{l}^{(M)} \quad (A2)$$

$$\mathbf{q}^{(M)}(\mathbf{l},n+1) = \sum_{L=n}^{M-1} \mathbf{l}^{(M-L)} \mathbf{q}^{(L)}(\mathbf{l},n) \quad (A3)$$

$$\theta^{(M)}(\lambda,n) = 0 \quad \text{if } n > M \quad (A4)$$

Firstly, we prove a preparative formula

$$\sum_k {}^{(S)}H'_{fk} b_{ki}^{(N)}(\lambda) = (\lambda^{(0)} - E_f) b_{fi}^{(N+1)}(\lambda) + \sum_{M=1}^{N+1} \lambda^{(M)} b_{fi}^{(N+1-M)}(\lambda) \quad (A5)$$

The stationary state equation is

$$\mathbf{l} c_f(\mathbf{l}) = \sum_k (E_k d_{fk} + {}^{(S)}H'_{fk}) c_k(\mathbf{l}) \quad (A6)$$

Its perturbed approximate equations are

$$(\mathbf{l}^{(0)} - E_j) c_j^{(0)}(\mathbf{l}) = 0 \quad (A7)$$

$$(\mathbf{l}^{(0)} - E_j) c_j^{(N)}(\mathbf{l}) + \sum_{L=1}^{N} \mathbf{l}^{(L)} c_j^{(N-L)}(\mathbf{l}) = \sum_k {}^{(S)}H'_{jk} c_k^{(N-1)}(\mathbf{l}) \quad (A8)$$

So we get

$$\sum_k {}^{(S)}H'_{fk} b_{ki}^{(N)}(\mathbf{l}) = \sum_k {}^{(S)}H'_{fk} \sum_{L=0}^{N} c_k^{(L)}(\mathbf{l}) c_i^{(N-L)}(\mathbf{l})^*$$

$$= \sum_{L=0}^{N} (\mathbf{l}^{(0)} - E_f) c_f^{(L+1)}(\mathbf{l}) c_i^{(N-L)}(\mathbf{l})^* + \sum_{L=0}^{N} \sum_{M=1}^{L+1} \mathbf{l}^{(M)} c_f^{(L+1-M)}(\mathbf{l}) c_i^{(N-L)}(\mathbf{l})^* \quad (A9)$$

We obtain that

the first term of (A9) $= (\mathbf{l}^{(0)} - E_f) \sum_{L=0}^{N+1} c_f^{(L)}(\mathbf{l}) c_i^{(N+1-L)}(\mathbf{l})^* = (\mathbf{l}^{(0)} - E_f) b_{fi}^{(N+1)}(\mathbf{l})$ (A10)

because of
$$(\lambda^{(0)} - E_f)c_f^{(0)}(\lambda)\Big|_{\lambda^{(0)}=E_l} = (\lambda^{(0)} - E_f)\delta_{fl} = 0 \tag{A11}$$

The second term of (A9) is relative to $\sum_{L=0}^{N}\sum_{M=1}^{L+1}$ . The region of two sums is in Fig. 2.

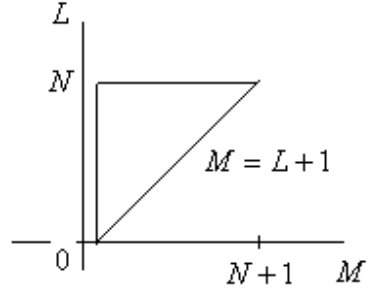

Fig. 2 The region of two sums is on the triangle (containing the boundary ).

We may exchange two sums by this way
$$\sum_{L=0}^{N}\sum_{M=1}^{L+1} = \sum_{M=1}^{N+1}\sum_{L=M-1}^{N}$$

and obtain that

the second term of (A9) $= \sum_{M=1}^{N+1}\lambda^{(M)}\sum_{L=M-1}^{N}c_f^{(L+1-M)}(\lambda)c_i^{(N-L)}(\lambda)^* = \sum_{M=1}^{N+1}\lambda^{(M)}b_{fi}^{(N-M+1)}(\lambda)$

(A12)

Then the relation
$$\sum_k {}^{(S)}H'_{fk}b_{ki}^{(N)}(\lambda) = (\lambda^{(0)} - E_f)b_{fi}^{(N+1)}(\lambda) + \sum_{M=1}^{N+1}\lambda^{(M)}b_{fi}^{(N+1-M)}(\lambda) \tag{A5}$$

is correct.

Now we use the inductive method to prove the time-dependent perturbed approximate solution $a_f^{(N)}(t)$ for arbitrary N is

$$a_f^{(N)}(t) = \sum_\lambda \left[b_{fi}^{(N)}(\lambda) + \sum_{n=1}^{N}\frac{1}{n!}\left(\frac{-it}{\hbar}\right)^n \sum_{M=n}^{N}b_{fi}^{(N-M)}(\lambda)\theta^{(M)}(\lambda,n)\right] \times \exp\left[\frac{i(E_f - \lambda^{(0)})t}{\hbar}\right]$$

(12)

If $N = 0$, (12) may be simplified as

$$a_f^{(0)}(t) = \sum_\lambda b_{fi}^{(0)}(\lambda) \exp\left[\frac{i(E_f - \lambda^{(0)})t}{\hbar}\right] = \sum_l \delta_{fl}\delta_{li} \exp\left[\frac{i(E_f - E_l)t}{\hbar}\right] = \delta_{fi} \quad (A13)$$

It is the solution of zeroth order time-dependent perturbed equation and satisfies the initial condition. If $N = 1$, (12) may be simplified as

$$a_f^{(1)}(t) = \sum_l \left[b_{fi}^{(1)}(l) + b_{fi}^{(0)}(l)q^{(1)}(l,1)(-it/\hbar)\right] \exp\left[\frac{i(E_f - l)t}{\hbar}\right]$$

$$= \sum_l \left[c_f^{(1)}(l)c_i^{(0)}(l)^* + c_f^{(0)}(l)c_i^{(1)}(l)^* + {}^{(S)}H'_{ll} c_f^{(0)}(l)c_i^{(0)}(l)^*(-it/\hbar)\right] \exp\left[\frac{i(E_f - E_l)t}{\hbar}\right]$$

$$= \begin{cases} \dfrac{{}^{(S)}H'_{fi}}{E_i - E_f}\left\{\exp\left[\dfrac{i(E_f - E_i)t}{\hbar}\right] - 1\right\} & f \neq i \\ {}^{(S)}H'_{ll}(-it/\hbar) & f = i \end{cases}$$

(A14)

It is the solution of the first order time-dependent perturbed equation and satisfies the initial condition.

Supposing that the solution of N order perturbed equation is (12), we shall prove that the solution of N+1 order perturbed equation has the form same as (12).

$$\sum_k {}^{(S)}H'_{fk} \exp\left[\frac{i(E_f - E_k)t}{\hbar}\right] a_k^{(N)}(t)$$

$$= \sum_l \left[\sum_k {}^{(S)}H'_{fk} b_{ki}^{(N)}(l) + \sum_{n=1}^N \frac{1}{n!}\left(\frac{-it}{\hbar}\right)^n \sum_{K=0}^{N-M} \left(\sum_k {}^{(S)}H'_{fk} b_{ki}^{(K)}(l)\right) q^{(N-K)}(l,n)\right]$$

$$\times \exp\left[\frac{i(E_f - l^{(0)})t}{\hbar}\right]$$

$$= \sum_l (l^{(0)} - E_f)\left[b_{fi}^{(N+1)}(l) + \sum_{n=1}^N \frac{1}{n!}\left(\frac{-it}{\hbar}\right)^n \sum_{K=0}^{N-n} b_{fi}^{(K+1)}(l)q^{(N-K)}(l,n)\right]$$

$$\times \exp\left[\frac{i(E_f - l^{(0)})t}{\hbar}\right]$$

$$+ \sum_l \left[\sum_{M=1}^{N+1} l^{(M)} b_{fi}^{(N-M+1)}(l) + \sum_{n=1}^N \frac{1}{n!}\left(\frac{-it}{\hbar}\right)^n \sum_{K=0}^{N-n} \sum_{M=1}^{K+1} l^{(M)} b_{fi}^{(K+1-M)}(l) q^{(N-K)}(l,n)\right]$$

$$\times \exp\left[\frac{i(E_f - l^{(0)})t}{\hbar}\right]$$

(A15)

The first term of (A15) =

$$\sum_{I} (I^{(0)} - E_f) \left[ b_{fi}^{(N+1)}(I) + \sum_{n=1}^{N+1} \frac{1}{n!} \left( \frac{-it}{\hbar} \right)^n \sum_{K=0}^{N+1-n} b_{fi}^{(K)}(I) q^{(N+1-K)}(I,n) \right]$$
$$\times \exp\left[ \frac{i(E_f - I^{(0)})t}{\hbar} \right]$$

A factor in second term of (A15) is

$$\sum_{K=0}^{N-n} \sum_{M=1}^{K+1} \lambda^{(M)} b_{fi}^{(K+1-M)}(\lambda) \theta^{(N-K)}(\lambda, n) = \sum_{K=0}^{N-n} \sum_{L=0}^{K} b_{fi}^{(L)}(\lambda) \lambda^{(K+1-L)} \theta^{(N-K)}(\lambda, n)$$
$$= \sum_{L=0}^{N-n} b_{fi}^{(L)}(\lambda) \theta^{(N+1-L)}(\lambda, n+1)$$

Therefore,
the second term of (A15)

$$= \sum_{I} \left[ \sum_{M=1}^{N+1} q^{(M)}(I,1) b_{fi}^{(N-M+1)}(I) + \sum_{n=1}^{N} \frac{1}{n!} \left( \frac{-it}{\hbar} \right)^n \sum_{L=0}^{N-n} b_{fi}^{(L)}(I) q^{(N+1-L)}(I, n+1) \right]$$
$$\times \exp\left[ \frac{i(E_f - I^{(0)})t}{\hbar} \right]$$

Substituting above results into (A15), we get

$$\sum_{k} H'_{fk} a_k^{(N)}(t) \exp\left[ \frac{i(E_f - E_k)t}{\hbar} \right] =$$
$$= \sum_{I} (I^{(0)} - E_f) \left[ b_{fi}^{(N+1)}(I) + \sum_{n=1}^{N+1} \frac{1}{n!} \left( \frac{-it}{\hbar} \right)^n \sum_{K=0}^{N+1-n} b_{fi}^{(K)}(I) q^{(N+1-K)}(I,n) \right]$$
$$\times \exp\left[ \frac{i(E_f - I^{(0)})t}{\hbar} \right] \qquad (A16)$$
$$+ \sum_{I} \left[ \sum_{K=0}^{N} b_{fi}^{(K)}(I) q^{(N+1-K)}(I,1) + \sum_{n=1}^{N} \frac{1}{n!} \left( \frac{-it}{\hbar} \right)^n \sum_{K=0}^{N-n} b_{fi}^{(K)}(I) q^{(N+1-K)}(I, n+1) \right]$$
$$\times \exp\left[ \frac{i(E_f - I^{(0)})t}{\hbar} \right]$$

Now let us prove that the expression of $a_f^{(N+1)}(t)$ is

$$a_f^{(N+1)}(t) = \sum_{I} \left[ b_{fi}^{(N+1)}(I) + \sum_{n=1}^{N+1} \frac{1}{n!} \left( \frac{-it}{\hbar} \right)^n \sum_{M=n}^{N+1} b_{fi}^{(N+1-M)}(I) q^{(M)}(I,n) \right]$$
$$\times \exp\left[ \frac{i(E_f - I^{(0)})t}{\hbar} \right] \qquad (A17)$$

In fact,

$$ih\frac{da_f^{(N+1)}(t)}{dt}$$

$$=\sum_\lambda (\lambda^{(0)} - E_f)\left[b_{fi}^{(N+1)}(\lambda) + \sum_{n=1}^{N+1} \frac{1}{n!}\left(\frac{-it}{\hbar}\right)^n \sum_{K=0}^{N+1-n} b_{fi}^{(K)}(\lambda)\theta^{(N+1-K)}(\lambda,n)\right]$$

$$\times \exp\left[\frac{i(E_f - \lambda^{(0)})t}{\hbar}\right]$$

$$+\sum_\lambda \left[\sum_{K=0}^{N} b_{fi}^{(K)}(\lambda)\theta^{(N+1-K)}(\lambda,1) + \sum_{n=1}^{N} \frac{1}{n!}\left(\frac{-it}{\hbar}\right)^n \sum_{K=0}^{N-n} b_{fi}^{(K)}(\lambda)\theta^{(N+1-K)}(\lambda,n+1)\right]$$

$$\times \exp\left[\frac{i(E_f - \lambda^{(0)})t}{\hbar}\right]$$

$$=\sum_k {}^{(S)}H'_{fk} \exp\left[\frac{i(E_f - E_k)t}{\hbar}\right]a_k^{(N)}(t)$$

(A18)

So (A17) satisfies the N+1 order perturbed approximate equation. In addition,

$$a_f^{(N+1)}(t=0) = \sum_\lambda b_{fi}^{(N+1)}(\lambda) = \sum_\lambda \sum_{L=0}^{N+1} c_f^{(L)}(\lambda)c_i^{(N+1-L)}(\lambda) = 0 \qquad (A19)$$

Therefore $a_f^{(N+1)}(t)$ satisfies the initial condition. So we prove that, if (12) is correct for N, then it must be correct for (N+1). We have prove this expression is correct for N=0,1. so it must be correct for arbitrary N.

## Appendix B: The expression of exact solution

According to the time-dependent perturbed approximate method, the exact solution is defined as

$$a_f(t) = \sum_{N=0}^{\infty} a_f^{(N)}(t)$$

$$= \sum_{\mathbf{l}} \left[\sum_{N=0}^{\infty} b_{fi}^{(N)}(\mathbf{l}) + \sum_{N=0}^{\infty} \sum_{M=1}^{N} b_{fi}^{(N-M)}(\mathbf{l})\left(\sum_{n=1}^{M} \frac{1}{n!}\left(\frac{-it}{\hbar}\right)^n q^{(M)}(\mathbf{l},n)\right)\right] \qquad (B1)$$

$$\times \exp\left[\frac{i(E_f - \mathbf{l}^{(0)})t}{\hbar}\right]$$

Now the task is to calculate the sum in this expression. Suppose

$$A = \sum_{K=0}^{\infty} A^{(K)}, \qquad B = \sum_{M=1}^{\infty} B^{(M)}; \qquad (B2)$$

then we have

$$A \cdot B = \left[ \sum_{K=0}^{\infty} A^{(K)} \right] \cdot \left[ \sum_{M=0}^{\infty} B^{(M)} \right]$$

$$= A^{(0)} B^{(1)} + (A^{(1)} B^{(1)} + A^{(0)} B^{(2)}) + (A^{(2)} B^{(1)} + A^{(1)} B^{(2)} + A^{(0)} B^{(3)}) + \cdots \quad (B3)$$

$$= \sum_{N=1}^{\infty} \sum_{M=1}^{N} A^{(N-M)} B^{(M)}.$$

Using this relation, (B1) may be written as

$$a_f(t) = \sum_{\lambda} \left\{ \sum_{N=0}^{\infty} b_{fi}^{(N)}(\lambda) + \sum_{K=0}^{\infty} b_{fi}^{(K)}(\lambda) \sum_{M=1}^{\infty} \left[ \sum_{n=1}^{M} \frac{1}{n!} \left( \frac{-it}{\hbar} \right)^n \theta^{(M)}(\lambda, n) \right] \right\}$$

$$\times \exp\left[ \frac{i(E_f - \lambda^{(0)})t}{\hbar} \right]$$

$$= \sum_{\lambda} \left[ \sum_{N=0}^{\infty} b_{fi}^{(N)}(\lambda) \right] \left\{ 1 + \sum_{n=1}^{\infty} \left[ \frac{1}{n!} \left( \frac{-it}{\hbar} \right)^n \sum_{M=n}^{\infty} \theta^{(M)}(\lambda, n) \right] \right\} \exp\left[ \frac{i(E_f - \lambda^{(0)})t}{\hbar} \right]$$

(B4)

We prove that the relation:

$$\left( \sum_{M=1}^{\infty} \lambda^{(M)} \right)^n = \sum_{M=n}^{\infty} \theta^{(M)}(\lambda, n). \quad (B5)$$

is correct.

According to (A2), we get

$$\sum_{M=1}^{\infty} \boldsymbol{1}^{(M)} = \sum_{M=1}^{\infty} \boldsymbol{q}^{(M)}(\boldsymbol{1}, 1) \quad (B6)$$

so (B5) is correct for $n = 1$. Suppose that (B5) is correct for $n$, then we have

$$\left( \sum_{M=1}^{\infty} \boldsymbol{1}^{(M)} \right)^{n+1} = \left( \sum_{L=1}^{\infty} \boldsymbol{1}^{(L)} \right) \cdot \sum_{M=n}^{\infty} \boldsymbol{q}^{(M)}(\boldsymbol{1}, n)$$

$$= \sum_{K=n+1}^{\infty} \sum_{M=n}^{\infty} \boldsymbol{q}^{(M)}(\boldsymbol{1}, n) \boldsymbol{1}^{(K-M)} \quad (B7)$$

$$= \sum_{K=n+1}^{\infty} \boldsymbol{q}^{(K)}(\boldsymbol{1}, n+1)$$

Therefore (B5) is correct for $n+1$. The conclusion is that the relation (B5) is correct for arbitrary $n$.

According to (A1),

$$\sum_{M=0}^{\infty} b_{fi}^{(M)}(\boldsymbol{1}) = \sum_{M=0}^{\infty} \sum_{L=0}^{M} c_f^{(L)}(\boldsymbol{1}) c_i^{(M-L)}(\boldsymbol{1})^*$$

$$= \left( \sum_{L=0}^{\infty} c_f^{(L)}(\boldsymbol{1}) \right) \cdot \left( \sum_{M=0}^{\infty} c_i^{(M)}(\boldsymbol{1})^* \right) = c_f(\boldsymbol{1}) c_i(\boldsymbol{1})^*$$

(B8)

Substituting (B5) and (B8) into (B4), we finally obtain

$$a_f(t) = \sum_{\boldsymbol{I}} c_f(\boldsymbol{I}) c_i(\boldsymbol{I})^* \left\{ 1 + \sum_{n=1}^{\infty} \frac{1}{n!} \left[ \frac{-it}{\hbar} \left( \sum_{L=1}^{\infty} \boldsymbol{I}^{(L)} \right) \right]^n \right\} \cdot \exp\left[ \frac{i(E_f - \boldsymbol{I}^{(0)})t}{\hbar} \right]$$

$$= \sum_{\boldsymbol{I}} c_f(\boldsymbol{I}) c_i(\boldsymbol{I})^* \exp\left[ \frac{i(E_f - \boldsymbol{I})t}{\hbar} \right]$$ (B9)

It is just the exact solution of Schrödinger equation. From this deducing process, we conversely prove that, using the stationary state perturbed method to expand $\boldsymbol{I}$, $c_f(\boldsymbol{I})$ in exact solution, we may obtain the different order of time-dependent solutions.

## Appendix C: Test the exact solution of Schrödinger equation

Now we test that, is

$$a_f(t) = \sum_{\boldsymbol{I}} c_f(\boldsymbol{I}) c_i(\boldsymbol{I})^* \exp\left[ \frac{i(E_f - \boldsymbol{I})t}{\hbar} \right]$$ (C1)

the exact solution of Schrödinger equation? Does it satisfy the initial condition? In fact,

$$i\hbar \frac{da_f(t)}{dt} = \sum_{\boldsymbol{I}} (\boldsymbol{I} - E_f) c_f(\boldsymbol{I}) c_i(\boldsymbol{I})^* \exp\left[ \frac{i(E_f - \boldsymbol{I})t}{\hbar} \right]$$

$$= \sum_k {}^{(S)}H'_{fk} \exp\left[ \frac{i(E_f - E_k)t}{\hbar} \right] \sum_{\boldsymbol{I}} c_k(\boldsymbol{I}) c_i(\boldsymbol{I})^* \exp\left[ \frac{i(E_k - \boldsymbol{I})t}{\hbar} \right]$$ (C2)

$$= \sum_k {}^{(S)}H'_{fk} \exp\left[ \frac{i(E_f - E_k)t}{\hbar} \right] a_k(t)$$

Therefore (C1) satisfies the Schrödinger equation. In addition, if $t = 0$

$$a_f(t=0) = \sum_{\boldsymbol{I}} c_f(\boldsymbol{I}) c_i(\boldsymbol{I})^* = \boldsymbol{d}_{fi}$$ (C3)

It is the special initial condition.

Welcome to [www.new.com](www.new.com)   Email:junhaoz@pub.shantou.gd.cn